\documentclass[aps,prl,twocolumn,showpacs,preprintnumbers,amsmath,amssymb,longbibliography]{revtex4-1}
\usepackage{hyperref}
\usepackage{bm}
\usepackage{graphics}
\usepackage{braket}
\usepackage{mathtools}
\usepackage{mathrsfs}
\usepackage{comment}
\usepackage{autobreak}
\usepackage{color}

\edef\restoreparindent{\parindent=\the\parindent\relax}
\usepackage{parskip}
\restoreparindent

\def\d{{\rm d}}

\def\CD{{\cal D}}

\def\d{\mathrm{d}}
\def\bz{\bar{z}}

\begin{document}

\title{Three-dimensional de Sitter holography and bulk correlators at late time}
\preprint{YITP-22-37}

\author{Heng-Yu Chen$^{a, b}$ and Yasuaki Hikida$^c$}

\affiliation{$^a$Department of Physics, National Taiwan University, Taipei 10617, Taiwan}
\affiliation{$^b$Physics Division, National Center for Theoretical Sciences, Taipei 10617, Taiwan}
\affiliation{$^c$Center for Gravitational Physics and Quantum Information, Yukawa Institute for Theoretical Physics, Kyoto University, Kyoto 606-8502, Japan}

%\date{\today}

\begin{abstract}

We propose an explicitly calculable example of holography on 3-dimensional de Sitter space by providing a prescription to analytic continue a higher-spin holography on 3-dimensional anti-de Sitter space. Applying the de Sitter holography, we explicitly compute bulk correlation functions on 3-dimensional de Sitter space at late time in a higher-spin gravity. These expressions are consistent with recent analysis based on bulk Feynman diagrams. Our explicit computations reveal how holographic computations could provide fruitful information.

\end{abstract}

\maketitle

%%%%%%%%  document begins  %%%%%%%%%

\section{Introduction}
Quantum gravity on de Sitter (dS) background is important to understand our early universe. In particular, the late time correlation functions on dS have been argued to give a clue to understand what happened at inflation era \cite{Maldacena:2002vr,Maldacena:2011nz,Arkani-Hamed:2015bza,Arkani-Hamed:2018kmz}. It is expected that dS/CFT correspondence \cite{Strominger:2001pn,Witten:2001kn,Maldacena:2002vr} is useful to formulate quantum gravity on dS as the better explored AdS/CFT correspondence \cite{Maldacena:1997re} does for quantum gravity on anti-de Sitter (AdS) space.
Until recently,  however, there has been only one concrete explicit prototype of dS/CFT correspondence, which uses higher-spin gravity on 4-dimensional dS \cite{Anninos:2011ui}.
The duality can be understood as an analytic continuation of higher-spin AdS$_4$ holography by \cite{Klebanov:2002ja}, see \cite{Sundborg:2000wp,Mikhailov:2002bp,Witten,Sezgin:2002rt} for previous works.
Recently, another explicit proposal was made on 3-dimensional de Sitter holography \cite{Hikida:2021ese,Hikida:2022ltr}, see \cite{Ouyang:2011fs} for a previous attempt. The purpose of this letter is to compute bulk correlation functions on 3-dimensional dS at late time by elaborating the proposal furthermore. To our best knowledge, this is the first computation of bulk dS correlators at late time based on an explicit concrete holographic setup. We compare our results with the recent analysis based on direct bulk Feynman diagram computations \cite{Sleight:2020obc,Sleight:2021plv}. See also \cite{Shukla:2016bnu,DiPietro:2021sjt} as well. Since lower dimensional examples are much tractable, we expect to learn a lot about mysterious dS/CFT correspondence through them.

For our purpose, we develop holographic method for bulk computations and higher-spin dS$_3$ holography. We first identify the 
phases relating the dS boundary operators with their AdS counterparts due to the analytic continuation from AdS.
We then elaborate the dS$_3$/CFT$_2$ correspondence in \cite{Hikida:2021ese,Hikida:2022ltr}.
In order to analyze bulk correlators, we consider an analytic continuation of Gaberdiel-Gopakumar duality \cite{Gaberdiel:2010pz}, which is between 3d Prokushkin-Vasiliev theory \cite{Prokushkin:1998bq} and 2d coset model
\begin{align}
    \frac{\text{SU}(N)_k \times \text{SU}(N)_1}{\text{SU}(N)_{k+1}} \, .\label{coset}
\end{align}
This coset was proven to describe W$_N$-minimal model \cite{Arakawa:2018iyk}.
We provide a prescription to perform an analytic continuation, which is different from the previous one
\footnote{In \cite{Hikida:2021ese,Hikida:2022ltr} a limit of $k \to -N$ is considered, while in this letter an 't Hooft limit is taken with large $|N|,|k|$ but finite $\lambda = N/(N+k)$}.
We then apply the Maldacena's holographic prescription to compute bulk dS correlators at late time given in \cite{Maldacena:2002vr} and find explicit expressions for 2- and 3-pt.\,correlators. We also examine a simple 4-pt.\,correlator.
We finally compare our results to generic arguments on bulk Feynman diagrams in \cite{Sleight:2020obc,Sleight:2021plv} and comment on the advantage of holographic approach.

\section{Preliminary}

We start by reviewing the AdS/CFT correspondence in order to explain how dS/CFT correspondence is obtained from an analytic continuation. We consider a gravity theory on $d+1$-dimensional AdS and $d$-dimensional CFT on the AdS boundary. 
In the bulk theory, we assume there exist symmetric tensor fields $\sigma_{i_1 \cdots i_s}^\text{AdS}$ with mass $m$ and spin $s$, and the scalar fields correspond to the ones with $s=0$. The CFT operators dual to the bulk fields are denoted by $J^{i_1 \cdots  i_s}_\text{AdS}$. For the Euclidean AdS metric, we use the Poincar\'e coordinates,
\begin{align}
\d s^2  = \ell_\text{AdS}^2\, \frac{\d y^2 + \d \vec{x} ^2}{y^2} \label{AdSmetric}
\end{align}
with the AdS radius $\ell_\text{AdS}$. 
We consider the region with $y \geq 0$ and the boundary is located at $y=0$. The bulk field behaves near $y=0$ as
\begin{align}
        \sigma_{i_1 \cdots i_s} ^\text{AdS}(y, \vec x) \sim \sigma_{i_1 \cdots  i_s}^{+,\text{AdS}} (\vec x)\, y ^{\Delta_+ - s} + \sigma_{i_1 \cdots  i_s}^{-,\text{AdS}} (\vec x)\, y ^{ \Delta_- - s}
\end{align}
with
\begin{align}
\ell_\text{AdS}^2\, m^2 = - (\Delta_+ \Delta_- + s) \, , \qquad \Delta_- = d -  \Delta_+  \ .
\end{align}
We use the flat boundary metric as $\d s^2 = \d \vec x ^2$, and the coupling between the bulk field and its dual operator is
\begin{align}
\ell^d_\text{AdS} \int \d^d \vec x\, \sigma^{\pm,\text{AdS}}_{i_1 \cdots i_s}\, J^{i_1 \cdots i_s}_{\pm,\text{AdS}} \, ,  \label{coupling}
\end{align}
where $\ell_\text{AdS}$-dependent factor is explicitly written.

We then consider dS/CFT correspondence.
The Poincar\'e patch for Lorentzian dS$_{d+1}$ is described by
\begin{align}
 \d s^2  = \ell ^2\,\frac{- \d \eta^2 + \d \vec{x} ^2}{\eta^2} \label{dSmetric}
\end{align}
with dS radius $\ell$. We consider the region $\eta \leq 0$ and a boundary is located at the future infinity $\eta \to - 0$.
The metric can be related to \eqref{AdSmetric} by
\begin{align}
y = - i \eta \ , \qquad \ell_\text{AdS} = - i \ell \, . \label{map}
\end{align}
As in the case of AdS$_{d+1}$, we consider bulk tensor fields on dS$_{d+1}$ denoted by $\sigma_{i_1 \cdots i_s}$, which are dual to boundary operators $J^{i_1 \cdots i_s}$.
In particular, the conformal dimension of the dual operator is
\begin{align}
\ell^2\, m^2 = \Delta_+ \Delta_- + s  \, , \qquad \Delta_- = d -  \Delta_+ \ .
\end{align}
In this letter, we consider the case with small $\ell^2 m^2$ 
in order to avoid subtlety associated with complex $\Delta_\pm$.
Near $\eta \to - 0$, the bulk field behaves as
\begin{align}
\begin{aligned}
    &\sigma_{i_1 \cdots i_s} (\eta, \vec x) \\
    &\sim  \sigma^+_{i_1 \cdots  i_s}  (\vec x)\, (-\eta) ^{\Delta_+ - s} + \sigma^-_{i_1 \cdots  i_s}  (\vec x)\, (-\eta) ^{ \Delta_- - s}  \, . \label{sphase}
\end{aligned}
\end{align}
Following \cite{Maldacena:2002vr}, we may identify $\sigma_{i_1 \cdots i_s}^\text{AdS}$ with $\sigma_{i_1 \cdots i_s}$ by generalizing the map \eqref{map} as follows.
Notice that bulk indices are lowered and raised by the bulk metric $g_{\mu \nu} \sim \ell^{2}$ but the boundary metric is independent of $\ell$, see, e.g., \cite{Anninos:2011ui}. It is thus natural to assign 
\begin{align}
    \ell^{-s}\, \sigma^\pm_{i_1 \cdots  i_s}  (\vec x) = i^{ \Delta_\pm - s }\, \ell_\text{AdS}^{- s}\, \sigma^{\pm,\text{AdS}}_{i_1 \cdots  i_s}  (\vec x)  \, , 
 \end{align}
and we have
\begin{align}
\begin{aligned}
    &\sigma^{\pm}_{i_1 \cdots  i_s}  (\vec x) = e^{i \frac{\pi}{2}\Delta_\pm}\, \sigma^{\pm , \text{AdS}}_{i_1 \cdots  i_s}  (\vec x)  \, , \\
    &J^\pm_{i_1 \cdots  i_s}  (\vec x) = e^{i \frac{\pi}{2}(d- \Delta_\pm)}\, J^{\pm , \text{AdS}}_{i_1 \cdots  i_s}  (\vec x) \, .
    \end{aligned}
\end{align}
For conserved currents with $\Delta_+ = s + d -2$, symmetric traceless tensors are
\begin{align}
    J^+_{i_1 \cdots  i_s}  (\vec x) = e^{i\frac{\pi}{2} (2 -s)}\, J^{+ , \text{AdS}}_{i_1 \cdots  i_s}  (\vec x) \, . \label{jreld}
\end{align}
In particular the energy momentum tensor with $s=2$ does not receive any phase factor and the standard operator product expansions are preserved.

As mentioned above, we would like to compute bulk correlators on dS at late time by following the prescription of \cite{Maldacena:2002vr}, which may be summarized as
\begin{align}
\Psi [\sigma^{\pm}_{i_1 \cdots i_s} ] = \left \langle \exp \left( \ell^d \int \d^d \vec x\, \sigma^{\pm}_{i_1 \cdots i_s}\, J^{i_1 \cdots i_s}_{\pm} \right) \right\rangle \, .  \label{HHWF}
\end{align}
The left hand side is the wave functional of universe for a fixed $d$-dimensional metric $g_{\mu \nu} = h_{\mu \nu}$  at late time $\eta \to -0$. 
The right hand side is computed by a certain CFT with operators $J^{i_1 \cdots i_s}_{\pm}$ coupled with their dual bulk fields $\sigma^{\pm}_{i_1\dots i_s}$.
The bulk correlation functions are then computed as expectation values as
\begin{align}
\left \langle\, \prod_{j=1}^m \psi_{j} (\vec x_j)\, \right \rangle = \int [\CD \psi_i]\, \left|\Psi \left[\psi_l\right]\right|^2\, \prod_{j=1}^m  \psi_{j} (\vec x_j) \, ,
 \label{wff}
\end{align}
where we set $\psi_j = \sigma^{\pm}_{i_1 \cdots i_{s_j}}$. 
See appendix A and, e.g.,  \cite{Arkani-Hamed:2018kmz} for more details.

\section{Higher-spin \texorpdfstring{AdS$_3$}{AdS3} holography}
In this letter,
we construct an explicit example of dS$_3$/CFT$_2$ correspondence by suitable analytic continuation of Gaberdiel-Gopakumar duality \cite{Gaberdiel:2010pz} as in \cite{Hikida:2021ese,Hikida:2022ltr}.
The gravity side of this
duality is given by the Prokushkin-Vasiliev theory \cite{Prokushkin:1998bq} on AdS$_3$, which consists of higher-spin gauge fields and two complex scalar fields. The higher-spin sector can be described by Chern-Simons gauge fields $A =A_\mu \d x^\mu, \bar A = \bar A_\mu \d x^\mu$. The gauge fields take values in higher-spin algebra hs$[\hat \lambda]$. The algebra can be defined such as to be truncated to sl$(N')$ at $\hat \lambda = N'$ with positive integer $N'$ even though we use $0 < \hat \lambda < 1$.
The Chern-Simons action is given by 
\footnote{We use the definition of Tr for the infinite dimensional algebra given in \cite{Vasiliev:1989re,Gaberdiel:2011wb}.}
\begin{align}
\begin{aligned}
 &S = S_\text{CS} [A] - S_\text{CS} [\bar A] \, , \\
 &S_\text{CS} = \frac{k_\text{CS}}{4 \pi} \int \text{Tr} \left( A \wedge \d A + \frac{2}{3} A \wedge A \wedge A \right) \, .
\end{aligned}
 \label{CSaction}
\end{align}
Here $k_\text{CS}$ is the level of the Chern-Simons theory, and it relates to gravity parameters as $k_\text{CS} =\ell_\text{AdS}/(4G_N)$, where $G_N$ is the Newton constant.
The asymptotic symmetry near the AdS boundary is found to be a W-algebra \cite{Henneaux:2010xg,Campoleoni:2010zq}. The central charge is the same as the Brown-Henneaux one \cite{Brown:1986nw} as
\begin{align} \label{BHcentral}
    c = 6 k_\text{CS} =  \frac{3 \ell_\text{AdS}}{2 G_N} \, .
\end{align}
Note that the classical gravity limit with $G_N \to 0$ corresponds to the large central charge limit $c \to \infty$.
More precisely speaking, the symmetry algebra is W$_\infty [\hat \lambda]$, which can be truncated to W$_{N'}$-algebra with spin-$s \, (=2,3,\ldots,N')$ currents at $\hat \lambda = N'$ \cite{Gaberdiel:2012ku}.
The two complex scalar fields $\phi_\pm$ have masses given by $\ell ^2_\text{AdS}\, m^2  =  \hat \lambda ^2 - 1 $.
%where we assign $0 < \hat \lambda < 1$. 

The CFT dual to the higher-spin theory on AdS$_3$ was proposed to be the W$_N$-minimal model at the 't Hooft limit \cite{Gaberdiel:2010pz}.  The W$_N$-minimal model can be described by a coset CFT \eqref{coset}
with the central charge
\begin{align}
c = (N-1) \left(1 - \frac{N(N+1)}{(N+k) (N+k+1)} \right) \, . \label{central}
\end{align}
The classical limit of higher-spin gravity is proposed to be dual to the 't Hooft limit, where $N,k \to \infty$ but  't Hooft parameter $\lambda = N/(N+k)$
is kept finite. Note that $0 < \lambda < 1$ for real positive integer $N,k$.
The proposal of \cite{Gaberdiel:2010pz} is that $\lambda$ is identified with $\hat \lambda$ appearing in the higher-spin theory. In the following, we only use  $\lambda$ to express the parameter.

The dual operators are given by two complex scalar operators $ \mathcal{O}_\pm^\text{AdS}(z,\bz)\, (\equiv\mathcal{O}^\text{AdS}_{\pm}(z) )$ and  holomorphic conserved spin-$s$ currents $J_{(s)}^\text{AdS}(z)$ with $\Delta_+ = s=2,3,\ldots$ (and anti-holomorphic currents $\bar J_{(s)}^\text{AdS}(\bz)$).
The conformal dimensions of $\mathcal{O}_\pm^\text{AdS}(z)$
are $2 h_\pm= 1 \pm \lambda$ at the 't Hooft limit. 
The 3-pt.\,functions of scalar-scalar-higher-spin current are computed from the bulk Vasiliev equations in \cite{Ammon:2011ua} as 
\begin{align}
\begin{aligned}
   & \langle\, \mathcal{O}_\pm ^\text{AdS}(z_1)\, \bar{ \mathcal{O}}_\pm ^\text{AdS}(z_2)\, J_{(s)}^\text{AdS} (z_3)\, \rangle  \\
    &= C_\pm^{(s)} 
    \left ( \frac{z_{12}}{z_{13} z_{23}} \right) ^s\,  \langle \, \mathcal{O}_\pm ^\text{AdS}(z_1) \, \bar{\mathcal{O}}_\pm ^\text{AdS}(z_2) \, \rangle
    \end{aligned}
    \label{3pt}
\end{align}
with
\begin{align}\label{3pt-coeff}
C_\pm^{(s)} = \frac{\eta^s_\pm}{2 \pi} \, \frac{\Gamma (s)^2}{\Gamma (2 s-1)} \frac{\Gamma (s \pm \lambda) }{ \Gamma (1 \pm \lambda)}
\end{align}
at the leading order in $1/c$.
The phase factors can be chosen arbitrary but here we set $\eta^s_+ = 1$ and $\eta^s_- = (-1)^s$ as in \cite{Hikida:2017byl}.
The holomorphic higher-spin currents are normalized as
\begin{align}
\begin{aligned}
   & \langle\, J_{(s)}^\text{AdS} (z)\, J_{(s)}^\text{AdS} (0)\, \rangle 
        =
        \frac{c B^{(s)}}{z^{2s}} \, , \\
    & B^{(s)} 
        =  \frac{1}{2 ^{2s} \pi^\frac{5}{2}}\, \frac{\sin (\pi \lambda)}{ \lambda (1 - \lambda^2)} \,
    \frac{\Gamma (s) \, \Gamma (s - \lambda)\, \Gamma (s + \lambda)}{ \Gamma \left(s - \frac12\right)} \, . 
\end{aligned}
    \label{2pt}
\end{align}
Since the higher-spin theory has one dimension-less parameter $k_\text{CS} \, (=c/6)$, the coupling to scalar fields is also organized by the same parameter. Thus the canonical normalization of 2-pt.\,functions of scalar operators is 
\begin{align} \label{OOk}
 \langle\, \mathcal{O}_\pm ^\text{AdS}(z_1)\, \bar{\mathcal{O}}_\pm ^\text{AdS}(z_2) \, \rangle \sim \frac{c}{|z_{12}|^{4 h_\pm}}
\end{align}
up to an overall real factor as shown in \cite{Anninos:2011ui}.

\begin{figure}
  \centering
  \includegraphics[width=8.6cm]{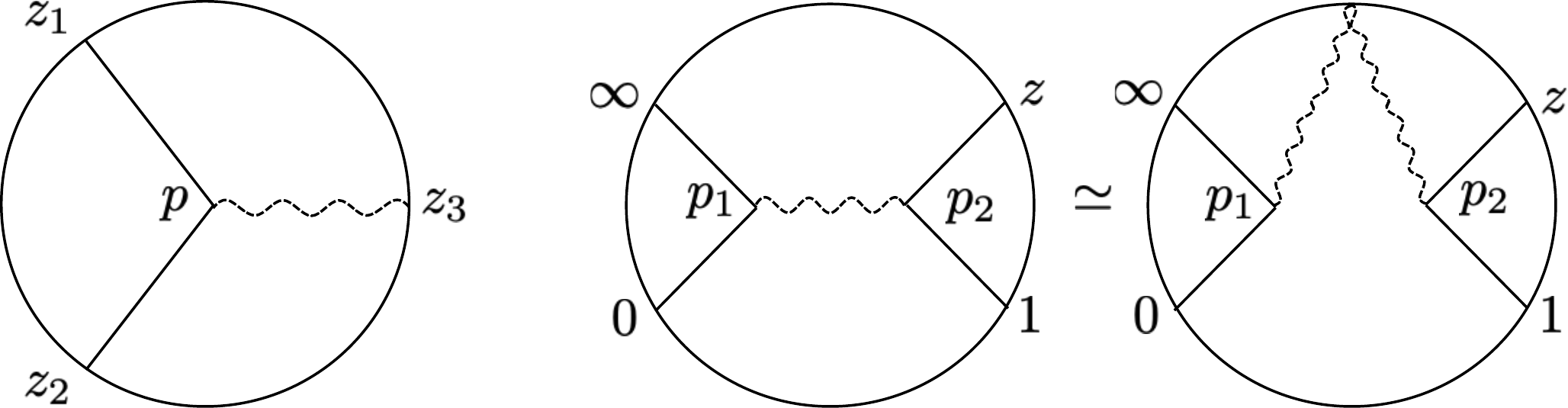}
 \caption{Witten diagrams for 3- and 4-pt.\,functions. The exchange diagrams may be expressed as products of 3-pt. functions via split representation of bulk-to-bulk propagators. }
  \label{fig:Witten}
\end{figure}

Here we remark that the 3-pt.\,correlators \eqref{3pt} can also be obtained from the bulk 3-pt.\,Witten diagram:
\begin{align}
    \int \d ^3 p K_{2 h_\pm,0} (z_1;p) K_{2 h_\pm,0} (z_2;p) K_{s,s} (z_3;p) \, , \label{Witten3t}
\end{align}
see fig.\,\ref{fig:Witten}. Here we represent a bulk point by $p=(y,z)$ and a bulk-to-boundary spin-$s$ propagator with dual dimension $\Delta$ by $K_{\Delta,s} (z'; p)$. While the $z_i$-dependence was fixed kinematically, this integral was evaluated in \cite{Costa:2014kfa} to produce the overall dynamical factor, and we can show that it matches with \eqref{3pt-coeff} up to an $s$-independent factor.

\section{Analytic continuation to \texorpdfstring{dS$_3$/CFT$_2$}{dS3/CFT2}}

We would like to map the bulk fields on AdS$_3$ to those on dS$_3$ applying \eqref{map}.
Due to the map, the level of Chern-Simons theory in \eqref{CSaction} has to be changed as \cite{Witten:1988hc}
\begin{align}
k_\text{CS} =  \frac{\ell_\text{AdS}}{4G_N} =  - i  \frac{\ell}{4G_N} = - i \kappa
\end{align}
with $\kappa \in \mathbb{R}$.
The symmetry at late time is supposed to be $W_\infty [\lambda]$-algebra with pure imaginary $c = - i \,  6 \kappa$ \cite{Strominger:2001pn,Ouyang:2011fs}, see \eqref{BHcentral}.
The masses of two complex scalars $\phi_\pm$ become $\ell ^2 m^2  = 1 - \lambda ^2 $.
Here we again assign $0 < \lambda < 1$ but now for $\ell^2  m^2$ to be positive.

The dual CFT is given by \eqref{coset} but the central charge \eqref{central} is now pure imaginary. 
We would like to provide a concrete prescription to perform an analytic continuation in the coset model. The symmetry underlying the holography is $W_\infty[\lambda]$-algebra, which was shown to be uniquely determined with two finite parameters $\lambda,c$ \cite{Gaberdiel:2012ku,Linshaw:2017tvv}.
The symmetry of the coset is realized by a special choice of $\lambda,c$ in terms of positive integer $k,N$.
These facts imply that the correct analytic continuation is keeping $\lambda$ fixed but setting $c = -  i c^{(g)}$ with $c^{(g)} > 0$ 
\footnote{We use the notation $c = - i c^{(g)}$ by following the previous works \cite{Hikida:2021ese,Hikida:2022ltr}.}.
Here we compute coefficient functions in the wave functional of universe via the relation \eqref{HHWF}, i.e.,  in terms of coset CFT \eqref{coset}, which can be also described by Toda CFT, see
\cite{Creutzig:2021ykz}.
We then change the parameters from $N,k$ to $c,\lambda$,
which can be easily analytically continued.

The CFT has two complex scalar operators $\mathcal{O}_\pm(z , \bar z ) (\equiv \mathcal{O}_\pm(z ))$ with conformal dimensions $2 h_\pm = 1 \pm \lambda$ at the 't Hooft limit and conserved higher-spin currents $J_{(s)} (z)$ (and its anti-holomorphic partner $\bar J_{(s)} (\bar z)$) with $s=2,3,\ldots$. The relation to bulk fields on AdS$_3$ can be read off from \eqref{sphase}.
For the scalar operator, we assign
\begin{align}
\mathcal{O}_\pm= - e^{ -  i  \pi h_\pm}  \mathcal{O}_\pm^\text{AdS} \, , \quad
\tilde{\mathcal{O}}_\pm =  - e^{ -  i  \pi h_\pm} \bar{\mathcal{O}}_\pm^\text{AdS} \, . \label{omap}
\end{align}
While $\bar{\mathcal{O}}_\pm^\text{AdS}$ is the complex conjugate of $\mathcal{O}_\pm^\text{AdS}$, the equation \eqref{omap} imply that $\mathcal{O}_\pm $ and $\tilde{\mathcal{O}}_\pm$ are no longer complex conjugate to each other. For higher-spin current, we have
\begin{align}
J_{(s)}  = - e^{-  \frac{i}{2} \pi s} J_{(s)}^\text{AdS} \, , \quad
\tilde J_{(s)} = - e^{- \frac{i}{2} \pi s } \bar J_{(s)}^\text{AdS} \, .
\label{jmap}
\end{align}

We then compute bulk correlators at late time from \eqref{wff}. The 2-pt.\,correlators can be computed as
\begin{align}
\langle\,  \phi_\pm (z_1)\, \tilde \phi _\pm (z_2)\, \rangle
      = - \frac{1}{2 \text{Re} \langle \mathcal{O}_\pm (z_1) \tilde{ \mathcal{O}}_\pm (z_2 ) \rangle} \, .
\end{align}
Note that only real part appears due to the square of wave functional as in \eqref{wff} and $\tilde \phi_\pm$ are complex conjugate of $\phi_\pm$.
Here the functional inverse is defined via (see, e.g., \cite{Dolan:2011dv,Osborn:2012vt}) 
\begin{align}
&k_{h,\bar h}\frac{1}{\pi}\int \d ^2 y\,\frac{(x-y)^{2h-2}(\bar x - \bar y)^{2 \bar h-2}}{(y -z)^{2h}(\bar y - \bar z)^{2\bar h}} = \delta^{(2)} (x - z) \, , \nonumber \\
&k_{h , \bar h} =  \frac{\Gamma (2 -2 \bar h)}{\Gamma (2 h -1)} = (-1)^{2 (h - \bar h)} \frac{\Gamma (2 -2h)}{\Gamma (2 \bar h -1)} \, . 
\label{K}
\end{align}
Noticing \eqref{OOk} and \eqref{omap}, we have
\begin{align}
    \begin{aligned}
       & \langle\, \mathcal{O}_\pm (z_1)\, \tilde{\mathcal{O}}_\pm (z_2) \, \rangle = - i  \, e^{- 2 i \pi h_\pm  }\, \frac{1}{|z_{12}|^{4 h_\pm}}
    \end{aligned}
\end{align}
by changing the overall factors of operators.
 From the formula \eqref{K}, we have
\begin{align}
\langle\,  \phi_\pm (z_1)\, \tilde \phi _\pm (z_2)\, \rangle
        &    = - a_{h_\pm} \frac{1}{\pi}\frac{\Gamma(2- 2 h_\pm)}{\Gamma(2 h_\pm -1)} \frac{1}{|z_{12}|^{4-4 h_\pm}\,} \label{2ptphi}
\end{align}
with
\begin{align}
            a_{h_\pm} = \frac{1}{2 \sin (2 \pi h_\pm)} \, . \label{aDelta}
\end{align}
The factor $a_{h_\pm}$ arises due to the analytic continuation and indeed it reproduces the corresponding factor $c_{\Delta}^\text{dS-AdS}$ with $\Delta = 2 h_\pm$ given in (2.15) of \cite{Sleight:2021plv}. 
Note that the factor  $c_{\Delta}^\text{dS-AdS}$ can be used also for spinning fields with conformal dimension $\Delta$ and spin $s$.

We then consider the higher-spin gauge field and describe the deviation from the background values by $\mu^{(s)}$ with $s=2,3,\ldots$.
Using the relation \eqref{jmap}, the 2-pt. coefficient functions are written as
\begin{align}
\begin{aligned}
&\langle\, J_{(s)} (z_1)\, J_{(s)} (z_2)\, \rangle %\\
= (-1)^{s+1}\frac{i c^{(g)}}{c}\,  \, 
\langle \, J_{(s)}^\text{AdS} (z_1)\, J_{(s)}^\text{AdS} (z_2)\, \rangle \, .
\end{aligned}
\end{align}
Thus the bulk correlators of higher spin fields at late time can be written as
\begin{align}
\begin{aligned}
   & \langle \mu_{(s)} (z_1) \mu_{(s)} (z_2) \rangle \\
   & \quad = a_{(s)} \frac{1}{ \pi \Gamma(2 s -1) c^{(g)} B^{(s)}}  \frac{1}{(z_{12})^{2-2s} (\bar z_{12})^{2}}\, .
\end{aligned} \label{2ptmu}
\end{align}
The factor
\begin{align} \label{as}
a_{(s)} = (-1)^{s+1}\frac{i}{2}
\end{align}
arises due to the analytic continuation as in the scalar case. However, the phase factor \eqref{as} does not match with $c_{\Delta}^\text{dS-AdS}$ for a massive spin-$s$ field in \cite{Sleight:2021plv} as it diverges for higher-spin gauge fields with $\Delta = s$. This implies that the massless limit of higher-spin field is quite subtle. Let us see an example of massive spin-1 field on Lorentzian dS$_{d+1}$ space-time. Its bulk two-point function was computed in (3.18)-(3.20) of \cite{Allen:1985wd}, but it diverges in the massless limit $m \to 0$. In the limit, gauge symmetry appears and analysis has to be redone from the beginning as in section 4 of \cite{Allen:1985wd}. In \cite{Sleight:2021plv}, a factor was inserted such that bulk massive spin-$s$ propagator on dS$_{d+1}$ has correct behavior near short distance. To obtain \eqref{as} by their method, we have to work directly with massless higher-spin gauge field or take a massless limit with a special care.

We then examine 3-pt.\,correlators at late time.
Using \eqref{omap} and \eqref{jmap}, the 3-pt.\,coefficient functions are related to \eqref{3pt} as
\begin{align}
\begin{aligned}
&\langle \, \mathcal{O}_\pm (z_1) \, \tilde{ \mathcal{O}}_\pm (z_2) \, J_{(s)}(z_3)\, \rangle \\
& \quad = i \, e^{ - i \frac{\pi}{2}(4 h_\pm  + s)} C_\pm^{(s)} \left( \frac{z_{12}}{z_{13}z_{23}}\right)^s \frac{1}{|z_{12}|^{4 h_\pm}}
\end{aligned} \label{ooj}
\end{align}
and similarly for $\langle  \mathcal{O}_\pm   \tilde{ \mathcal{O}}_\pm  \tilde J_{(s)} \rangle$.
The sum of \eqref{ooj} and the conjugate of $\langle  \mathcal{O}_\pm   \tilde{ \mathcal{O}}_\pm  \tilde J_{(s)} \rangle$ becomes
\begin{align}
 - 2 \sin \left( (4 h_\pm +s) \frac{\pi}{2} \right) C_\pm^{(s)} \left( \frac{ z_{12}}{ z_{13} z_{23}}\right)^s \frac{1}{|z_{12}|^{4 h_\pm}} \, .
\end{align}
In order to read off the bulk 3-pt.\,correlators, we need to multiply two of \eqref{2ptphi} and one of \eqref{2ptmu} and integrate over the positions, see \eqref{m3pt}.
We then obtain
\begin{align}
\begin{aligned} \label{bulk3pt}
&\langle\,  \phi_\pm (z_1)\, \tilde \phi _\pm (z_2)\, \mu_{(s)} (z_3) \rangle \\
&\quad= - \lambda_{h_\pm,h_\pm,s} \frac{\Gamma (\mp \lambda) \Gamma (1 \mp \lambda)}{\Gamma(s)^2 \Gamma (1 -s \pm \lambda) \Gamma(s \pm \lambda )} \\
& \quad \times \frac{C_\pm^{(s)}}{c^{(g)} B^{(s)}} \left( \frac{ z_{12}}{ z_{13} z_{23}}\right)^{1-s} \left( \frac{\bar z_{12}}{ \bar z_{13} \bar z_{23}}\right) \frac{1}{|z_{12}|^{4 h_\mp}} 
\end{aligned}
\end{align}
with
\begin{align}
   \lambda_{h_\pm,h_\pm,s} 
   = 2  a_{h_\pm}^2 a_{(s)} \sin \left( (4 h_\pm +s) \frac{\pi}{2}  \right) \, .
\end{align}
The same factor appears in (3.24) of \cite{Sleight:2021plv} up to the subtlety associated to the massless limit mentioned above.

\section{Four-point correlators}
In the coset model \eqref{coset}, the 4-pt.\,functions of scalar operators 
were computed in \cite{Papadodimas:2011pf} exactly in all orders of $N,k$.
In this letter, we analyze a simple one given by
\begin{align}
\begin{aligned}
 G_{-+}(z) &= \langle\, \mathcal{O}^\text{AdS}_- (\infty) \, \mathcal{O}^\text{AdS}_+ (1)\, \bar{\mathcal{O}}^\text{AdS}_+ (z)\, \bar{\mathcal{O}}^\text{AdS}_- (0 )\, \rangle  \\
&= |1 - z|^{- 2 \Delta_+} |z|^{\frac{2}{N} } \left| 1 + \frac{1 -z}{Nz}\right|^2 \, .
\end{aligned} \label{Gmp}
\end{align}
Following the analytic continuation procedures explained above, we should be able to obtain 4-pt.\,bulk correlator
\begin{align} \label{4ptdS}
&\langle\, \phi_- (\infty)\,  \phi_+ (1)\, \tilde \phi_+ (z)\, \tilde \phi_- (0 )\,\rangle \, 
\end{align}
in dS$_3$ at late time. 

The expansion of $G_{-+}(z)$ in \eqref{Gmp} in terms of global conformal blocks was given in \cite{Hikida:2017byl} as 
\begin{align}
\label{G-+exps0}
&G_{- +} (z) = |1 - z|^{- 2{\left(1+\lambda-\frac{1-\lambda^2}{c}\right)}} + |1 - z|^{-2{ (1+\lambda)}} \\ &\times \left[\sum_{s=2}^\infty (-1)^s \frac{C^{(s)}_- C^{(s)}_+}{ c B^{(s)}}\,(1 - z)^s {}_2F_1 \left(s,s; 2s ; 1- z\right) + \text{c.c.} \right] \nonumber
\end{align}
up to the order $1/c^2$. 
In particular, the conformal dimension of $\mathcal{O}_+$ can be expanded as 
$\Delta_+ = 1 + \lambda - (1 - \lambda^2)/c + \mathcal{O}(c^{-2})$.
The expansion now can be written as a sum over holomorphic and anti-holomorphic conserved spin-$s$ exchange with $s=2,3,\ldots , \infty$ and the coefficients are given as products of 3-pt.\,functions.
In terms of Witten diagram, the 4-pt.\,function should be computed from
\begin{align}
    \label{Witten4t}
   & \int \d ^3 p_1 \d ^3 p_2
    K_{2 h_-,0} (\infty;p_1) K_{2 h_+,0} (0;p_1) \\
   & \quad\times G_{s,s}(p_1;p_2) K_{2 h_+,0} (z;p_2) K_{2 h_-,0} (1;p_2)  \, , \nonumber
\end{align}
see fig.\,\ref{fig:Witten}.
Here $G_{\Delta,s}(p_1;p_2)$ represents the bulk-to-bulk spin-$s$ propagator with dual dimension $\Delta$. 
In the split representation, it can be expressed as 
\begin{align} \label{split}
   & G_{\Delta,s}(p_1;p_2) = \int_{-\infty}^\infty \frac{\d \nu}{\nu^2 + (\Delta -1)^2} \Omega_{\nu,s}(p_1;p_2) \, , \\
    &\Omega_{\nu , s} (p_1 ; p_2)= \frac{\nu^2}{\pi}\int \d ^2 z K_{1 + i \nu ,s} (p_1 ,z) K_{1 - i \nu ,s} (p_2 ,z) \nonumber
\end{align}
up to contact term contributions \cite{Costa:2014kfa}. Picking up poles in \eqref{split}, the 4-pt.\,function becomes a product of 3-pt. functions, and the integration over the boundary coordinates leads to a conformal partial wave
\begin{align} 
&\mathcal{I}_{s,0} (z ) 
 = z^s {}_2F_1 (s,s; 2s ; z)
+ \frac{\Gamma (2s -1)\Gamma(2s)}{\Gamma(s)^4} \nonumber\\
& \times z^{1-s} {}_2F_1 (1-s,1-s;2-2s;z) \bar z {}_2F_1 (1,1;2;\bar z)
\label{shadow}
\end{align}
up to an unimportant overall constant. See \eqref{Is0} and \cite{Simmons-Duffin:2012juh} for details.
In terms of conformal partial waves, the 4-pt.\,function can be expanded as:
\begin{align}
\label{G-+exps2}
&G_{- +} (z) = |1 - z|^{ - 2{\left(1+\lambda-\frac{1-\lambda^2}{c}\right)}}  + |1 - z|^{- 2(1+\lambda)} \\ &\times \left[ \sum_{s=2}^\infty (-1)^s \frac{C^{(s)}_- C^{(s)}_+}{ c B^{(s)}}\, \mathcal{I}_{s,0} (1-z ) - \frac{1-\lambda^2}{c} \ln \bar z  + \text{c.c.} \right] \, . \nonumber
\end{align}

As in \eqref{G-+exps2}, there are three kinds of terms in the expansions of 4-pt.\,function \eqref{Gmp},  each has a natural interpretation in the AdS and dS bulks. The first term is just the product of 2-pt.\,functions, corresponding to disconnected Witten diagrams. The second term is the sum of conformal partial waves, which can be expressed in lower-pt.\,coefficient functions as mentioned above. When we perform the path integral over $\phi_\pm , \bar \phi_\pm$, they will be over counted if we include them in the 4-pt.\,coefficient function. Thus, it is convenient to remove it beforehand. The third term corresponds to the contact 4-pt.\,interaction in the bulk. 
Analytically continuing from AdS to dS spaces, the connected parts of their respective 4-pt.\,CFT correlators can be related via:
\begin{align}
\begin{aligned}
&\langle\, \mathcal{O}_- (\infty) \, \mathcal{O}_+ (1)\, \tilde{\mathcal{O}}_+ (z)\, \tilde{\mathcal{O}}_- (0 )\, \rangle _ \text{c} \\
&= - i \, 
\langle\, \mathcal{O}^\text{AdS}_- (\infty) \, \mathcal{O}^\text{AdS}_+ (1)\, \bar{\mathcal{O}}^\text{AdS}_+ (z)\, \bar{\mathcal{O}}^\text{AdS}_- (0 )\, \rangle_\text{ c }\, ,
\end{aligned}
 \label{OOOO}
\end{align}
where the subscript ``$\text{c}$'' indicates the connected part. 
Notice that while using \eqref{omap} yields a phase factor $e^{- i  \pi(2h_+ + 2h_-)} =1$, the overall $- i$ comes from the fact that the connected part is proportional to $c$ if we use the normalization \eqref{OOk} \footnote{We can show that any connected $n$-pt.\,function is proportional to $c$ with the normalization \eqref{OOk} by repeating operator product expansions to reduce it to products of 3-pt.\,functions. This is consistent with the bulk perturbation by $k_\text{CS}$.}. 
The contribution from the third term vanishes in this case after taking its real part.

Thus, the only non-trivial contribution comes from that corresponding to the dS bulk exchange diagrams, which can be written as
\begin{align}
&\langle\, \phi_- (\infty)\,  \phi_+ (1)\, \tilde \phi_+ (z)\, \tilde \phi_- (0 )\,\rangle_ \text{c} = \sum_{s=2}^\infty \frac{\lambda_{h_+,h_+,s}\lambda_{h_-,h_-,s}}{ a_{(s)} c^{(g)}}
 \nonumber \\
&\times  \frac{\Gamma(- \lambda)\Gamma(1 - \lambda) \Gamma(\lambda) \Gamma(1 + \lambda)}{\Gamma(1 - s + \lambda) \Gamma (s + \lambda)\Gamma(1 - s - \lambda) \Gamma (s - \lambda)}  \\
& \times\frac{1}{ |1 - z|^{2 (1 - \lambda)} }\left[ (-1)^s \frac{C^{(s)}_- C^{(s)}_+}{ B^{(s)}}\, \mathcal{I}_{s,0} (1-z )  + \text{c.c.} \right] \, . \nonumber
\end{align}
The factor $\lambda_{h_\pm,h_\pm,s}$ comes from that in \eqref{bulk3pt} and the division by $a_{(s)}$ eliminates the over counting of \eqref{2ptmu}.
The result is consistent with the generic expression, say, in (4.39) of \cite{Sleight:2021plv}.

\section{Discussion}
In this letter, we computed bulk dS$_3$ correlators at late time by developing holographic method and dS$_3$/CFT$_2$ correspondence. The expressions are consistent with the previous analysis of \cite{Sleight:2020obc,Sleight:2021plv} based on bulk Feynman diagrams in the in-in formulation, whose review may be found in \cite{Weinberg:2005vy}. Let us end by briefly commenting these two complementary approaches exploring the dS/CFT holography and where our results fit in here.

Notice that, in the in-in formulation, there are two types of interaction vertices with time-ordering and anti-time ordering and the corresponding propagators connecting them in the dS bulk.
This is related to consider both signs for analytic continuation instead of our prescription \eqref{map}, see \cite{Sleight:2020obc,Sleight:2021plv}.
Moreover, as in the case of AdS computations, bulk 4-pt.\,dS correlators can reduce to the evaluation of the product of two 3-pt.\,functions by applying a formula analogous to \eqref{split}, see, e.g., \cite{Sleight:2020obc,Sleight:2021plv}. 
However in such computations, further integrals over spectral parameter $\nu$ are needed to make full comparisons with CFT correlators, which are usually very difficult.

In holographic method, wave functional and its complex conjugation are involved as in \eqref{wff} and the integration over boundary fields provides connections between them. We can see how the two methods complement each other from our explicit calculations.
As an advantage of holographic method, there are no integrals over spectral parameter $\nu$ in \eqref{wff}, and the difficulty is avoided by directly evaluating the coefficient functions via dual CFT. 
In our explicit example, the non-trivial information is included in the the third term of \eqref{G-+exps2}. The contribution vanishes at the end of computation, but the corresponding term survives for different 4-pt.\,correlators obtained from $\langle  \mathcal{O}_\pm^\text{AdS}  \bar{ \mathcal{O}}_\pm^\text{AdS} \mathcal{O}_\pm^\text{AdS} \bar{ \mathcal{O}}_\pm^\text{AdS} \rangle$. 
See \cite{Papadodimas:2011pf,Hikida:2017byl} for their exact forms and conformal block expansions. 
In addition, coefficient functions can be obtained even with finite $\lambda,c$ from the explicit dual CFT. For these reasons, holographic method would enable us to evaluate more complicated bulk correlators at loop levels, on dS$_3$ cosmological backgrounds \cite{Krishnan:2013zya}, and so on.
We are planing to report on more details on the relation between two methods and further analysis of bulk correlators in near future.

\vspace{5mm}
%%%%%%%%%%%%%%%%%%%%%%%%%%%%%
{\bf Acknowledgements} 
%%%%%%%%%%%%%%%%%%%%%%%%%%%%%
We are grateful to Thomas Creutzig, Kenta Suzuki, Tadashi Takayanagi, Yusuke Taki, Seiji Terashima and Takahiro Uetoko for useful comments. We would like to particularly thank Tatsuma Nishioka for the collaboration at the early stage and other related work. The work of H.\,Y.\,C. is supported in part by Ministry of Science
and Technology (MOST) through the grant 110-2112-M-002-006-.
This work of Y.\,H. was supported by JSPS Grant-in-Aid for Scientific Research (B) No.\,19H01896, Grant-in-Aid for Scientific Research (A) No.\,21H04469, and Grant-in-Aid for Transformative Research Areas (A) ``Extreme Universe'' No.\,21H05187.

%\bibliography{dSCFT}


\appendix

\section{Appendix A: Bulk correlators from wave functional}
\label{app:wf}

In this appendix, we collect formulas to compute bulk correlators from wave functional of universe.
Expanding the wave functional \eqref{HHWF} by the fields $\psi_j$ at the late time, we may write
\begin{align}
&\Psi [\psi_l]  \label{exp}\\
&=\exp \left[  \sum_{m  \geq 2} \int \d^d \vec x_1 \cdots \d^d \vec x_{m}   C_{m} (\{ \vec x_l\}) \prod_{j=1}^m \psi_j (\vec x_j) \right]\, .  \nonumber
\end{align}
The proposal here is that the coefficient functions $C_{m} (\{ \vec x_l\})$ can be computed by certain dual conformal field theory. 

The explicit forms of correlation functions in terms of CFT correlators are given as follows (see, e.g., \cite{Maldacena:2002vr,Maldacena:2011nz,Arkani-Hamed:2018kmz} for their momentum space expressions).
The 2-pt.\,functions are written as
\begin{align}
    \begin{aligned}
       \langle\,  \sigma_{i_1  \cdots i_s} (\vec x_1)\, \sigma^{j_1 \cdots j_s} (\vec x_2)\, \rangle
            =
            - \frac{\Pi^{j_1 \cdots j_s}_{i_1 \cdots i_s} (\hat x_{12})}{2\, \text{Re} \, \langle\, J_s (\vec x_1)\, J_s (\vec x_2)\,  \rangle} \, , 
    \end{aligned}
\end{align}
where  we have used the expression
\begin{align}
    \langle\,  J_{i_1  \cdots i_s} (\vec x_1)\, J^{j_1 \cdots j_s} (\vec x_2)\, \rangle =  \Pi^{j_1 \cdots j_s}_{i_1 \cdots i_s} (\hat x_{12})\,  \langle\, J_s (\vec x_1)\, J_s (\vec x_2)\,  \rangle
\end{align}
with the unit vector $\hat x = \vec x /|\vec x|$.
The 3-pt.\,functions used in the main context are
\begin{align} 
   & \langle\, \phi_1 (\vec x_1)\, \phi_2  (\vec x_2)\, \sigma_{i_1 \cdots i_s} ( \vec x_3)\, \rangle  \label{m3pt}\\
    & \
        = - \int\prod_{l=1}^3 \d ^d \vec x_l '  \text{Re}\,\langle\,  \mathcal{O}_1 (\vec x_1 ')\, \mathcal{O}_2 (\vec x_2 ')\, J_{j_1 \cdots j_s} (\vec x_3 ')\,  \rangle \nonumber  \\
& \times 
        \frac{ \Pi^{j_1 \cdots j_s}_{i_1 \cdots i_s} (\vec x_{33'}) } { 4 \left[ \prod_{i=1}^2 \text{Re}\,  \langle\, \mathcal{O}_i (\vec x_i)\, \mathcal{O}_i (\vec x_i ')\,  \rangle  \right] \text{Re}\, \langle\, J_s (\vec x_3)\, J_s (\vec x_3 ')\,  \rangle } \ .  \nonumber
\end{align}
Here we set for $s=0$ as $\phi = \sigma_{i_1 \cdots i_s}$ and $\mathcal{O} = J^{i_1 \cdots i_s}$.
The scalar 4-pt.\,function can be decomposed as
\begin{align} 
\label{4pt}
\begin{aligned}
   & \langle\, \phi_1 (\vec x_1)\, \phi_2  (\vec x_2) \, \phi_3 (\vec x_3)\, \phi_4 (\vec x_4)\, \rangle  \\
 &       =\int\prod_{l=1}^4 \d ^d \vec x_l '
        \frac{ \langle \mathcal{O}^4 \rangle_\text{c} + \langle \mathcal{O} ^4\rangle _\text{d} } {  \prod_{i=1}^4 (-2\,   \text{Re} \, \langle\, \mathcal{O}_i (\vec x_i)\, \mathcal{O}_i (\vec x_i ') \, \rangle ) }
\end{aligned}
\end{align}
with
\begin{align}
  &  \langle\, \mathcal{O}^4 \,\rangle_\text{c} 
        =
        2\, \text{Re}\, \langle\,  \mathcal{O}_1 (\vec x_1 ')\, \mathcal{O}_2 (\vec x_2 ') \,  \mathcal{O}_3 (\vec x_3 ')\, \mathcal{O}_4 (\vec x_4 ')\, \rangle \, , \label{O4c}\\
  &  \langle\, \mathcal{O}^4 \,\rangle_\text{d}  \nonumber \\
        &=\sum_s \int \d ^d \vec y_1 \d ^d  \vec y_2 2\, \text{Re}\,\langle \, \mathcal{O}_1 (\vec x_1 ') \, \mathcal{O}_2 (\vec x_2 ')  , J_{j_1 \cdots j_s} (\vec y_1) \,  \rangle \nonumber \\
      & \times 
        \frac{\Pi^{j_1 \cdots j_s}_{i_1 \cdots i_s} (\vec y_{12}) \text{Re}  \, \langle \, J^{i_1 \cdots i_s} (\vec y_2 ) \, \mathcal{O}_3 (\vec x_3 ') \, \mathcal{O}_4 (\vec x_4 ') \, \rangle  } {  \text{Re} \, \langle\, J_s (\vec y_1) \, J_s ( \vec y_2) \, \rangle } \, .  \label{O4d} 
\end{align}
For $\langle \, \mathcal{O} ^4 \,\rangle  _d$, we have to sum up all intermediate operators.

\section{Appendix B: Conformal integrals}
\label{app:ci}

In order to extract information of bulk correlators from the coefficient functions of wave functional, we need to perform some conformal integrals. Here we summarize some formulas, see, e.g., \cite{Dolan:2011dv,Osborn:2012vt}.

For 2-pt.\,correlators, we have used the formula \eqref{K}. For 3-pt.\,correlators, it is convenient to apply the formula
\begin{align}
\begin{aligned}
&k_{h , \bar h} \frac{1}{\pi}\int \d ^2 y \frac{1}{(z -y)^{2 - 2 h} (\bar z - \bar y)^{2 - 2 \bar h} } \mathcal{F}^{h , \bar h}_{12} (y , \bar y) \\
&= \frac{\Gamma(1 - h - h_{12}) \Gamma(1 - \bar h + \bar h_{12})}{\Gamma (\bar h + \bar h_{12}) \Gamma (h - h_{12})}\mathcal{F}^{1-h , 1- \bar h}_{12} (y , \bar y)
\end{aligned}
\end{align}
with
\begin{align}
\begin{aligned}
&\mathcal{F}^{h , \bar h}_{12} (z , \bar z)
= \frac{1}{z_{12}^{h_1 + h_2 - h} (z_1 - z)^{h+h_{12}} (z_2 - z)^{h - h_{12}}} \\ &\quad \times \frac{1}{\bar z_{12}^{\bar h_1 + \bar h_2 - \bar h} (\bar z_1 - \bar z)^{\bar h+\bar h_{12}} (\bar z_2 - \bar z)^{\bar h - \bar h_{12}}} \, .
\end{aligned}
\label{calFhh}
\end{align}
Here we have used the notation such as $z_{12} = z_1 - z_2$, $h_{12} = h_1 - h_2$ and so on.

For 4-pt.\,correlator, we expand it in terms of conformal partial waves as in \eqref{G-+exps2}.
We define a shadow operator $\hat \Phi_{1-h, 1 - \bar h}$ by 
\begin{align}
\begin{aligned}
& \hat \Phi_{1-h,1- \bar h} (x)  \\
& = k_{h,\bar h }\frac{1}{\pi}
        \int \d ^2 y\, \frac{1}{(x - y)^{2 - 2h} (\bar x - \bar y)^{2 - 2 \bar h}}\, \Phi_{h, \bar h} (y) \, ,
\end{aligned}
\end{align}
where $k_{h , \bar h}$ was defined in \eqref{K}.
The conformal partial wave of spin-$s$ exchange can be obtained from integrals
\begin{align}
&\frac{k_{1-s,1} k_{s,0}}{\pi^2 c B^{(s)}}\int \d ^2 x\, \d ^2 y\, \langle\, \mathcal{O}_-^\text{AdS} (\infty)\, \bar{\mathcal{O}}_-^\text{AdS} (0)\, J_{(s)}^\text{AdS} (x)\, \rangle\, \nonumber \\
& \times \frac{1}{(x -y)^{2-2s} (\bar x - \bar y)^{2}}\, \langle\, J_{(s)}^\text{AdS} (y)\, \mathcal{O}_+^\text{AdS} (1)\, \bar{\mathcal{O}}_+^\text{AdS} (z) \, \rangle \nonumber \\
& = \frac{\Gamma(2s)}{\pi c B^{(s)}}\int \d ^2 x\, \langle\, \mathcal{O}_-^\text{AdS} (\infty) \, \bar{\mathcal{O}}_-^\text{AdS} (0) \, J_{(s)}^\text{AdS}  (x)\, \rangle\,  \nonumber \\
& \times \, \langle \, \hat J_{(s)}^\text{AdS}  (x)\, \mathcal{O}_+^\text{AdS} (1)\, \bar{\mathcal{O}}_+^\text{AdS} (z) \, \rangle \label{Is0}\\
& = (-1)^s  \frac{1}{|1 -z|^{2 \Delta_+}}\frac{C_-^{(s)} C_+^{(s)}}{c B^{(s)}} \mathcal{I}_{s,0} (1 -z) \, , \nonumber
\end{align}
where $\mathcal{I}_{s,0} (z)$ is given in \eqref{shadow},
see, e.g., \cite{Simmons-Duffin:2012juh}.

\end{document}